\documentclass[useAMS,usenatbib]{mn2e}
\usepackage{times}
\usepackage{verbatim}
\usepackage{amsmath}
\usepackage{graphicx}
\usepackage{wrapfig} 
\usepackage{caption}
\usepackage{hyperref} 

\usepackage{color}
\usepackage{aas_macros}
\usepackage{graphicx}
\usepackage{epsf}
\usepackage{graphics}
\usepackage{rotating}
\usepackage{amssymb}   
\usepackage{setspace}  
\usepackage{lscape}

\newcommand{\be}{\begin{equation}}
\newcommand{\ee}{\end{equation}}

\newcommand{\kms}{\,km\,s$^{-1}$}
\newcommand{\mpc}{$h^{-1}$Mpc}

\title{2MTF V. Cosmography, $\beta$, and the residual bulk flow}

\author[Springob et~al.]
{Christopher M. Springob$^{1,2}$, Tao~Hong~$^{3,1,2}$, Lister~Staveley-Smith~$^{1,2}$,
\newauthor
Karen~L.~Masters~$^{4,5}$, Lucas~M.~Macri~$^{6}$, 
B\"arbel~S.~Koribalski~$^{7}$, 
D.~Heath~Jones~$^{8}$,
\newauthor
Tom~H.~Jarrett~$^{9}$, Christina Magoulas~$^{9}$, Pirin Erdo{\u g}du$^{10}$\\
$^1$International Centre for Radio Astronomy Research, The University of
Western Australia, Crawley, WA 6009 Australia\\ 
$^2$ARC Centre of Excellence for All-sky Astrophysics (CAASTRO)\\
$^{3}$National Astronomical Observatories, Chinese Academy
  of Sciences, 20A Datun Road, Chaoyang District, Beijing 100012,
  China.\\
 $^{4}$Institute for Cosmology and Gravitation, University 
of Portsmouth, Dennis Sciama Building, Burnaby Road, Portsmouth 
PO1 3FX, UK\\
$^{5}$South East Physics Network (www.sepnet.ac.uk)\\
$^{6}$George P. and Cynthia Woods Mitchell Institute for Fundamental Physics and 
Astronomy, Department of Physics and Astronomy, Texas A\&M University, \\4242 TAMU, 
College Station, TX 77843, USA\\ 
$^{7}$CSIRO Astronomy \& Space Science, Australia Telescope National 
Facility, PO Box 76, Epping, NSW 1710, Australia\\ 
$^{8}$Department of Physics and Astronomy, Macquarie University, Sydney, NSW 2109, Australia\\ 
$^{9}$Astronomy Department, University of Cape Town, Private Bag X3. 
Rondebosch 7701, Republic of South Africa\\
$^{10}$Australian College of Kuwait, PO Box 1411, Safat 13015, Kuwait\\ 
}

\begin{document}

\maketitle

\begin{abstract}
Using the Tully-Fisher relation, we derive peculiar velocities for the 2MASS Tully-Fisher Survey and describe the velocity field of the nearby Universe.  We use adaptive kernel smoothing to map the velocity field, and compare it to reconstructions based on the redshift space galaxy distributions of the 2MASS Redshift Survey (2MRS) and the IRAS Point Source Catalog Redshift Survey (PSCz).  With a standard $\chi^2$ minimization fit to the models, we find that the PSCz model provides a better fit to the 2MTF velocity field data than does the 2MRS model, and provides a value of $\beta$ in greater agreement with literature values.  However, when we subtract away the monopole deviation in the velocity zeropoint between data and model, the 2MRS model also produces a value of $\beta$ in agreement with literature values.  We also calculate the `residual bulk flow': the component of the bulk flow not accounted for by the models.  This is $\sim 250$ \kms\ when performing the standard fit, but drops to $\sim 150$ \kms\ for both models when the aforementioned monopole offset between data and models is removed.  This smaller number is more in line with theoretical expectations, and suggests that the models largely account for the major structures in the nearby Universe responsible for the bulk velocity.\end{abstract}


\section{Introduction}

The velocity field of galaxies exhibits deviations from Hubble flow
due to inhomogeneities in the large-scale distribution of matter.
By studying the galaxy peculiar velocity field, we can explore the large-scale distribution of matter in the local Universe and so test
cosmological models and measure cosmological parameters.

With $c$ representing the speed of light and $v_{pec}$ representing a galaxy's peculiar velocity, we define the `peculiar redshift' $z_{pec}$ as the peculiar velocity in redshift units, given by
\be
z_{pec} = v_{pec}/c
\ee
where $z_{pec}$ is related to the observed redshift $z_{obs}$ and the redshift due to Hubble flow $z_H$ according to
\be
(1 + z_{obs}) = (1 + z_H)(1 + z_{pec})
\ee
as given by \citet{harrison74}.  The low redshift approximation of this relation is
\be v_{pec} \approx cz_{obs} - cz_{H} \approx cz_{obs} - H_0 D \ee
where $H_0$ is the Hubble constant and $D$ is the galaxy's comoving distance.

The measurement of peculiar velocities is thus intertwined
with the measurement of distances, in that redshift-independent
distances are needed in combination with redshifts to extract peculiar velocities.  Most of the largest peculiar velocity surveys have made use of one of two redshift-independent distance indicators: the Tully-Fisher relation (TF; \citealt{tully77}) and the Fundamental Plane relation (FP; \citealt{djorgovski87x}, \citealt{dressler87a}).  These scaling relations express the luminosity of a spiral galaxy as a power-law function of its rotational velocity and the radius of an elliptical galaxy as a power law function of its surface brightness and velocity dispersion respectively.

The earliest peculiar velocity surveys using these distance indicators, such as \citet{aaronson82} and \citet{lyndenbell88}, included no more than a few hundred galaxies.  Many of these surveys were concatenated into the Mark III catalog (\citealt{willick95}, \citealt{willick96}).  Among the earliest individual TF surveys to include more than $\sim 1000$ galaxies were a set of surveys conducted by Giovanelli, Haynes, and collaborators (e.g., \citealt{giovanelli94}, \citealt{giovanelli95}, \citealt{giovanelli97}, \citealt{haynes99a}, \citealt{haynes99b}).  These surveys were combined, along with additional data, to create the SFI++ survey (\citealt{masters06x}; \citealt{springob07}), which included TF data for $\sim 5000$ galaxies.

At the time of its release, SFI++ was the largest peculiar velocity survey compiled.  However, both its selection criteria and data sources were quite heterogeneous.  The 2MASS Tully-Fisher survey (2MTF) was envisioned as a `cleaner' all-sky TF survey, with more stringent selection criteria, drawn from a more homogeneous dataset, and extending to significantly lower Galactic latitudes.  It draws on galaxies selected from the 2MASS Redshift Survey (2MRS, \citealt{huchra12}), and uses photometry from the 2MASS Extended Source Catalogue \citep{jarrett00x} and spectroscopy from the Green Bank Telescope (GBT), Parkes radio telescope, Arecibo telescope, and other archival HI catalogs.  The archival data overlaps heavily with the archival dataset used for SFI++, and the HI width measurement procedure and peculiar velocity derivation also follows the SFI++ procedure closely.  The 2MTF template relation was presented by \citet{masters08}, while the first cosmological analysis using the dataset, a measurement of the bulk flow, was presented by \citet{hong14}.

In this paper, we examine the cosmography of the observed 2MTF velocity field, and compare it to two reconstructions of the predicted peculiar velocity field, which assume that the matter distribution traces the galaxy distribution.  The first is the Erdogdu et al. (submitted, updated from \citealt{erdogdu06}) reconstruction of the aforementioned 2MRS.  The second is the \citet{branchini01} reconstruction of the {\it IRAS} Point Source Catalog Redshift Survey (PSCz, \citealt{saunders00}).  In comparing to these reconstructions, we compute the $\chi^2$ agreement between the observed and predicted velocity fields, the value of the redshift space distortion parameter $\beta$, and the amplitude and direction of the `residual bulk flow', which represents the component of the bulk flow of the local universe not predicted by models.

The history of such comparisons between observed peculiar velocity fields and predictions made from large all-sky redshift surveys goes back to analyses such as those of \citet{kaiser91}, \citet{shaya92}, \citet{hudson94}, \citet{davis96}, and \citet{hudson04}.  More recently, we have seen the comparisons between SFI++ and 2MRS performed by \citet{davis11a}, between SFI++ and 2M++ \citep{lavaux11} performed by \citet{carrick15}, and between the 6dF Galaxy Survey (6dFGS) peculiar velocity field \citep{springob14} and both the 2MRS and PSCz reconstructions performed by \citet{magoulas12}, Magoulas et al. (in prep.), and \citet{springob14}.  These later papers find good overall agreement between the observed and predicted velocity fields, but disagree on whether the predicted fields can explain the amplitude of the bulk flow.  2MTF is well suited to examine this issue because of its all-sky coverage, sampling the sky down to Galactic latitudes of $|b|=5^{\circ}$.

The paper is arranged as follows: In Section 2, we describe the 2MTF observational dataset as well as the 2MRS and PSCz model velocity fields.  In Section 3, we describe the derivation of peculiar velocities.  In Section 4, we describe the cosmography of the 2MTF velocity field, present our comparison of the field to the model velocity fields, and discuss our results.  We summarize our results in Section 5.

\section{Data}

\subsection{2MTF TF data}

The 2MTF target list was compiled from the set of all 2MRS \citep{huchra12} spirals with total K-band magnitude $K_s<11.25$, CMB frame redshift $cz<10,000$ \kms, and axis ratio $b/a<0.5$.  There are $\sim 6000$ galaxies that meet these criteria, though many of them are quite faint in HI, and observationally expensive to observe with the available single dish radio telescopes.  Thus only a fraction of these objects are included in the final 2MTF sample.

We have combined archival data with observations made as part of the Arecibo Fast Legacy ALFA Survey \citep{giovanelli05}, and observations made with the Green Bank Telescope \citep{masters14} and Parkes Telescope \citep{hong13}.  While the new GBT and Parkes observations preferentially targeted late type spirals, the archival data includes all spiral types.  The complete 2MTF sample includes 2018 galaxies (note that this is separate from the 888 cluster galaxies used to fit the template TF relation in \citealt{masters08}).  However, as noted in Section 3.2, for the analysis performed in this paper, we exclude large outliers from the TF relation, which reduces the number of objects to 1985.

\subsubsection{Photometric Data}

All of our photometry is drawn from the 2MRS catalog.  2MRS is an all-sky redshift survey, consisting of $\sim 43,000$ redshifts of 2MASS galaxies, extending in magnitude to $K_s < 11.75$ and Galactic latitude $|b|>5^{\circ}$.  While the final sample \citep{huchra12} has a limiting K-band magnitude of 11.75, the 2MTF survey began before 2MRS was complete, and so the magnitude limit was set to the somewhat shallower value of 11.25.  For the analysis presented in this paper, we use total K-band magnitudes.

While the \citet{masters08} template used I-band and J-band axis ratios, we use 2MASS J/H/K co-added axis ratios (see \citealt{jarrett00x}).  As explained by \citet{hong14}, the dispersion between these two definitions of axis ratio is $\sim 0.096$, and we account for this in deriving the scatter of the TF relation, but find no systematic trend between the two definitions towards larger or smaller values.

Internal dust extinction and $k$-correction were done as described by \citet{masters08}, and updated in the erratum \citet{masters14b}.  The 2MRS total magnitudes are already corrected for Galactic extinction, so no further correction is necessary.

\subsubsection{Spectroscopic Data}

The primary source of archival data used here is the Cornell HI archive \citep{springob05}, which offers HI spectroscopy for $\sim 9000$ galaxies in the local universe, as observed by single dish radio telescopes.  We find 1038 galaxies from this dataset with high quality spectra that match the 2MTF selection criteria.  We also include HI data from \citet{theureau98}, \citet{theureau05}, \citet{theureau07}, and \citet{matthewson92}, as well as the Nancay observations from Table A.1 of \citet{paturel03}.  The raw observed HI widths drawn from these sources were then corrected for inclination, redshift stretch, instrumental effects, turbulence, and smoothing, according to the prescriptions of \citet{springob05}, which were then updated by \citet{hong13}.

To supplement the archival data, we made new observations using the Parkes and Green Bank telescopes between 2006 and 2012.  These datasets were presented by \citet{hong13} and \citet{masters14} respectively, but we briefly summarize them here.

The observations were divided up in such a way that Parkes was used to target galaxies in southern declinations ($\delta<-40^{\circ}$), while the GBT targeted more northern galaxies ($\delta>-40^{\circ}$), but outside the ALFALFA survey region.  (See Figure 1 for the sky distribution of objects.)  The GBT observations targeted 1193 galaxies in position switched mode, with the spectrometer set at 9 level sampling with 8192 channels.  After smoothing, the velocity resolution was 5.15 \kms .  727 galaxies were detected, with 483 of them being deemed sufficiently high quality detections to be included in our sample.

\begin{figure*} \centering 
\begin{minipage}{175mm}
\includegraphics[width=1.0\textwidth]{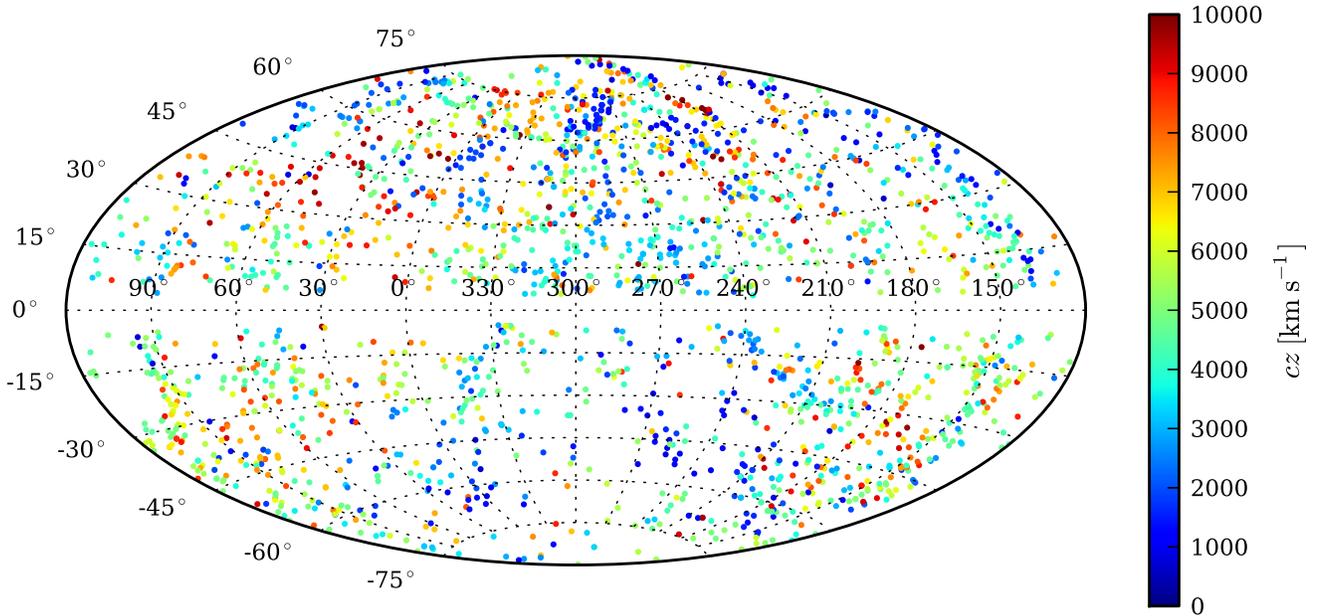} 
\caption{Distribution of 2MTF galaxies in Galactic latitude ($l$) and longitude ($b$), shown in an Aitoff projection.  Galaxies are color-coded by redshift, according to the colorbar on the right hand side of the plot.\label{FIG1}} 
\end{minipage} 
\end{figure*}

For the Parkes observations, we targeted 305 galaxies which did not already have high quality HI width measurements in the literature.  Of these, we obtained width measurements suitable for inclusion in our sample for 152 galaxies.  The multibeam correlator produced raw spectra with a velocity resolution of 1.6 \kms , though Hanning smoothing broadened the resolution to 3.3 \kms .  Each of the GBT and Parkes spectra were analyzed using the IDL routine {\it awv\_fit.pro}, which is based on the method used for the Cornell HI archive \citep{springob05}, which in turn is based on the earlier approach developed by Giovanelli, Haynes, and collaborators (e.g., \citealt{giovanelli97b}).  We use the $W_{F50}$ width algorithm, as defined in those papers.  It involves fitting a line to either side of the HI line profile, and measuring the width from the points on each line representing $50\%$ of the flux minus {\it rms} value.  This approach was also adopted for use in the ALFALFA survey.

In addition to the archival, GBT, and Parkes datasets, we also included ALFALFA data from the initial data release \citep{haynes11}.  The catalog presented in that data release covers roughly $40\%$ of the final survey.  However, the ALFALFA team has provided us with additional unpublished data from the survey, current as of October 2013.  In total then, we cover $\sim 66\%$ of the ALFALFA sky.  From this sample, we use 576 galaxy widths, which will be updated once the final release of ALFALFA is available.

As noted above, this gives us a final sample of 1985 galaxies when TF outliers are excluded.  The redshift histogram of this sample can be found in Figure 2.

\begin{figure}
\includegraphics[width=0.9\columnwidth]{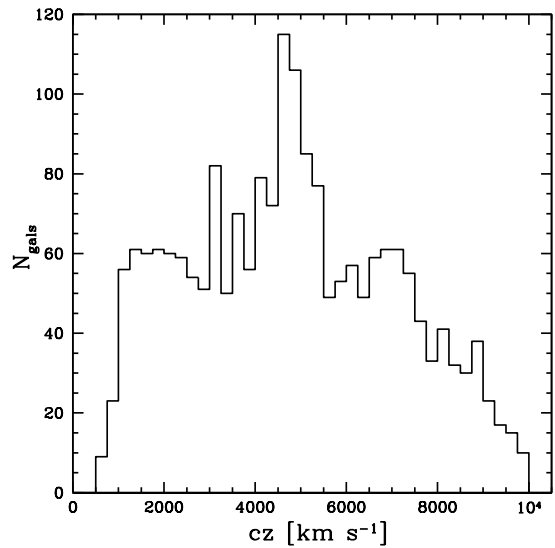}
\caption{Redshift distribution of galaxies in 2MTF in the CMB
  reference frame.  The bin width is 250 \kms.
\label{FIG2}}
\end{figure}

\subsection{PSCz model velocity field}

The IRAS Point Source Catalog Redshift Survey (PSCz, \citealt{saunders00}) includes 15,500 galaxies, covering $84\%$ of the sky, with most of the missing sky area lying at low Galactic latitudes.  \citet{branchini99} reconstructs the local density and velocity field from this survey, using a spherical harmonic expansion method proposed by \citet{nusser94}.  The method assumes a linear mapping between the PSCz spatial distribution of galaxies and the matter distribution, with an input assumed bias parameter $\beta = 0.5$.  The grid spacing of the model is 2.8 \mpc, extending to a distance of $180$ \mpc \ from the origin in each direction.

We convert the PSCz velocity grid from real space to redshift space, so that we may compare to the redshift space positions of the 2MTF galaxies.  Each gridpoint is assigned to its position in redshift space by adding its reconstructed peculiar velocity to its real distance.  The reconstructed velocities are then linearly interpolated onto a regularly spaced grid (again with 2.8 \mpc~ resolution) in redshift space.  The problem of triple-valued regions is mitigated in the original \citet{branchini99} reconstruction, by the authors collapsing galaxies within clusters, and applying a method devised by \citet{yahil91} to determine the locations of galaxies along those lines of sight.

Following the reconstruction of the PSCz velocities onto this redshift space grid, we assign line-of-sight predicted PSCz velocities to each of the galaxies in the 2MTF sample.  This is done by interpolating the nearest gridpoints to the position of the galaxy in question.  The interpolation is done with inverse distance weighting.  This method was also used by \citet{springob14} to match 6dFGS peculiar velocities to predictions from models.

\subsection{2MRS model velocity field}

We also make use of a density/velocity field reconstruction from 2MRS itself.  As noted in Section 2.1, the final 2MRS data release \citep{huchra12} includes redshifts for 44,699 galaxies, and covers $91\%$ of the sky (excluding only the Galactic zone of avoidance).  \citet{erdogdu06} reconstructed the density/velocity field from an earlier 2MRS data release \citep{huchra05}, and this has now been updated to make use of the final dataset (Erdo{\u g}du et~al.\, submitted).

The reconstruction method is described in detail by \citet{erdogdu06}.  It most closely follows the method used by \citet{fisher95}, again assuming that the matter distribution follows the galaxy distribution, with an assumed $\beta = 0.4$.  It involves decomposing the redshift space density field of galaxies into spherical harmonics, and smoothing with a Wiener filter.  The
reconstruction gives densities and velocities on a grid in supergalactic
cartesian coordinates with gridpoints spaced by 8 \mpc \ and extending
to a distance of 200 \mpc \ from the origin in each direction.

As with PSCz, we convert this grid from real space to redshift space.  The resulting redshift space grid in this case has grid spacing of 4 \mpc.  Again, as with PSCz, we then use interpolation between gridpoints to assign model velocities to galaxies in the 2MTF sample.

\section{Tully-Fisher distances and peculiar velocities}

We follow the same basic procedure that was followed for the SFI++ derivation of peculiar velocities by \citet{springob07}, which relied on the calibration of the TF relation from the template relation of \citet{masters06x}.  This in turn followed the procedure used by the SFI and SCI surveys (\citealt{giovanelli94}, \citealt{giovanelli95}, \citealt{giovanelli97b}).

In brief, we use the J-, H-, and K-band TF template relations from \citet{masters08} to derive the peculiar velocities of the individual galaxies in our sample.  The \citet{masters08} template sample includes 888 galaxies, while the sample presented in this paper includes 1985 galaxies which does not overlap with the template sample.  The template relations found by \citet{masters08} are
\begin{align}
&M_K - 5\log h = -22.188 - 10.74(\log W -2.5),
\notag \\
&M_H - 5\log h = -21.951 - 10.65(\log W -2.5),
\\
&M_J - 5\log h = -21.370 - 10.61(\log W -2.5),
\notag
\end{align}
where $W$ is the corrected HI width in units of \kms, and $M_K$, $M_H$, and $M_J$ are the corrected absolute magnitudes in the three bands.

One difference between the SFI++ approach and the one employed here is that we work with logarithmic distance ratios throughout, rather than converting to linear peculiar velocities.  This is done because the distance errors (as well as the individual errors on line width and magnitude) are approximately log-normal.  We thus make use of the quantity
\be
\Delta d^{*} = \log {\left(d_z \over d_{TF}^{*}\right)} = {-\Delta M \over 5}
\ee
where $\Delta M = M_{obs}-M(W)$ is the difference between the corrected absolute magnitude $M_{obs}$, calculated using the redshift distance of the galaxy ($d_z$), and the magnitude $M(W)$ derived from the TF template relation.  $d_{TF}^{*}$ is the distance to the galaxy derived from the Tully-Fisher relation, but not corrected for Malmquist/selection bias.  We then refer to $\Delta d^{*}$ as the logarithmic distance ratio (uncorrected for Malmquist bias).  In Section 3.2, we discuss the Malmquist bias correction, at which point this quantity is replaced by $\Delta d$, the logarithmic distance ratio (corrected for Malmquist bias).

As noted by \citet{hong14}, the fact that \citet{masters08} uses a different set of axial ratios means that we must derive new measurements of the intrinsic scatter in the TF relation.  \citet{hong14} does this, and arrives at the relations
\begin{align}
&\epsilon_{int,K}=0.44-0.66(\log W -2.5),
\notag \\
&\epsilon_{int,H}=0.44-0.95(\log W -2.5),
\\
&\epsilon_{int,J}=0.46-0.77(\log W -2.5),
\notag
\end{align}
for the scatter in the TF relation {\it in magnitude units}.  To convert these to logarithmic distance units, one must divide the $\epsilon_{int}$ values by 5.

\subsection{Galaxy groups}

\citet{crook07} describes a galaxy group catalog for 2MRS, derived using a `friends-of-friends' algorithm.  55 of these groups include more than one member in our sample.  To eliminate the effects of motions within galaxy groups, we thus fix the redshift used when calculating the logarithmic distance ratio in Equation 5 to the group redshift for all of our group galaxies.  However, the magnitude offset $\Delta M$ is still calculated separately for each galaxy.

\subsection{Selection bias}

`Malmquist bias' is the term used to describe biases originating from the interaction between the spatial distribution of objects and the selection effects \citep{malmquist24}.  It results from
the coupling between the random distance errors and the apparent
density distribution along the line of sight. There are two types of
distance errors to consider. The first is `inhomogeneous
Malmquist bias', which arises from local density variations
due to large-scale structure along the line of sight. This bias is most
pronounced when measuring galaxy distances in real space.
This is because the large distance errors scatter the measured galaxy distances away from overdense regions, creating artificially inflated
measurements of infall onto large structures. By contrast,
when the measurement is done in redshift space, the much smaller
redshift errors mean that this effect tends to be negligible (see e.g.,\citealt{strauss95}).

For the 2MTF sample, we
are measuring galaxy distances and peculiar velocities in redshift
space rather than real space. In this case, inhomogeneous
Malmquist bias is negligible, and the form of Malmquist bias that
we must deal with is of the second type, known as homogeneous
Malmquist bias, which affects all galaxies independently of their
position on the sky. It is a consequence of both (1) the volume effect,
which means that more volume is covered within a given solid
angle at larger distances than at smaller distances, and (2) the selection
effects, which cause galaxies of different luminosities, radii,
velocity dispersions etc., to be observed with diminishing completeness with increased distance. We note, however, that different
authors use somewhat different terminology, and the latter effect
described above is often simply described as `selection bias'.

The approach one takes in correcting for this bias depends in
part on the selection effects of the survey.  In our case, we used homogeneous criteria in determining
which galaxies to observe.  However, many of the galaxies that met our selection criteria yielded nondetections, marginal detections, or there was some other problem with the spectrum that precluded the galaxy's inclusion in our sample.  Our final sample, then, lacks the same homogeneity as the original target sample, and our selection bias correction procedure must account for this.

We adopt the following procedure:

1) Using the stepwise maximum likelihood method \citep{efstathiou88}, we derive the K-band luminosity function $\Phi (M_k)$, as a function of K-band absolute magnitude $M_k$, for
all galaxies in 2MRS that meet our K-band apparent magnitude,
Galactic latitude, morphological, and axis ratio criteria. For this
purpose, we include galaxies beyond the 10,000 \kms redshift limit,
to simplify the implementation of the luminosity function derivation.  The resulting luminosity function of this sample is found in Figure 3.  We fit a
Schechter function \citep{press74} to this distribution, and find parameters $M_k*=-23.1$ and $\alpha=-1.10$.  The final parameter of the Schechter function is the normalization.  The stepwise maximum likelihood method cannot derive this parameter, but it is not needed for our method in any case.  This luminosity function has a steeper slope than the 2MASS K-band luminosity function dervied by \citet{kochanek01}, who find $\alpha=-0.87$.  Our morphological selection criteria presumably contribute to this difference, though \citet{jones06} notes that there is some disagreement between the \citet{kochanek01} luminosity function and others.

\begin{figure}
\includegraphics[width=0.9\columnwidth]{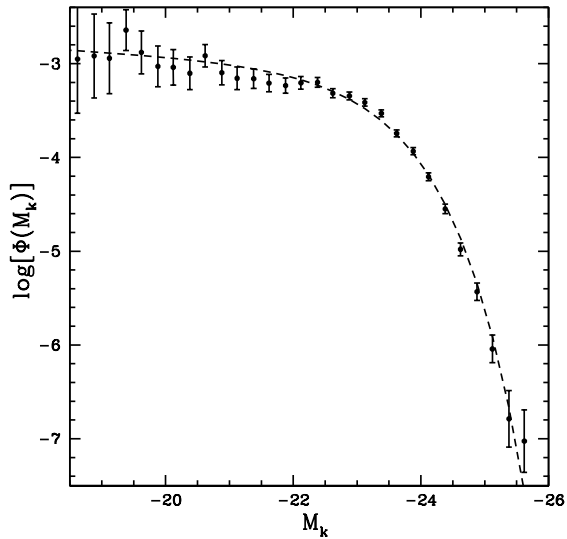}
\caption{K-band luminosity function $\Phi (M_k)$, as a function of absolute magnitude $M_k$, for 2MTF target galaxies, derived using the stepwise maximum likelihood method.  Errorbars represent the uncertainty from Poisson statistics.  The dashed line represents the bets fit Schechter function, which gives Schechter parameters $M_k*=-23.1$ and $\alpha=-1.10$.  The vertical axis is in dex units, but the normalization is arbitrary, as the stepwise maximum likelihood method is unable to fit the normalization.  However, the normalization is also irrelevant, as the correction for selection bias only requires us to know the {\it relative} luminosity function.
\label{FIG3}}
\end{figure}

2) We define the `completeness' as the fraction of the target sample that is included in our final catalog for a given apparent magnitude bin.  We divide up the sky between two regions: one covering all declinations north of $\delta = -40^{\circ}$, and the other covering all declinations south of $\delta = -40^{\circ}$.  The $\delta = -40^{\circ}$ boundary is the declination at which the sample transitions from GBT to Parkes observations.  After experimenting with further subdivisions of the sky, we conclude that the completeness as a function of apparent magnitude remains constant within each region, and assume that that holds in the analysis that follows.  We compute this completeness function separately for the two sky regions, by taking
the ratio of observed galaxies to galaxies in the target sample for
K-band apparent magnitude bins of width 0.25 mag.

3) For each of the galaxies in the sample, we compute the uncorrected logarithmic distance ratio ($\Delta d^*$) probability distribution, assuming a Gaussian distribution, with an uncorrected $1\sigma$ scatter of ${\epsilon_d}^*$.  For each logarithmic distance ratio value ($\Delta d_i$) within $2\sigma$ of the nominal value of $\Delta d^*$, we weight the probability by $w_i$, where $1/w_i$ is the completeness (from Step 2) integrated across the K-band luminosity function (from Step 1), evaluated at the $\Delta d_i$ in question.  Using the luminosity function in this manner involves converting it into apparent magnitudes, using the relevant distance modulus.

4) Finally, we fit a Gaussian to the re-weighted probability distribution.  This gives us the corrected logarithmic distance ratio ($\Delta d$), and its error ($\epsilon_d$).  We then cut from the sample any galaxies deemed outliers, defined as those for which the deviation from $\Delta d = 0$ cannot be accounted for by the quadrature sum of a 300 \kms scatter in peculiar velocities plus $3\sigma$ deviation from the TF relation.  43 galaxies are eliminated by this criterion, leaving us with a total sample of 1985 galaxies.

The $\Delta d$ histograms can be found in \citet{hong14}.  As noted in that paper, the mean TF distance error translates to $\sim 22\%$ in all three wavebands.

\section{Results and discussion}

\subsection{Velocity field cosmography}

Because of the significant distance errors obtained for each individual galaxy, we use adaptive kernel smoothing to get a cosmographic view of the velocity field.  We set up a 3D redshift space grid in supergalactic cartesian coordinates, with all gridpoints spaced 4\mpc\ apart.  At each gridpoint, we compute adaptively smoothed values of $\Delta d$ for both the PSCz reconstructed velocity field, and the observed 2MTF field.  This is done following a procedure outlined in \citet{springob14}, that draws on methods used by \citet{silverman86} and \citet{ebeling06}.  We summarize this method below.

We define $\Delta d({\mathbf r_i})$ as the logarithmic distance ratio at redshift space position ${\mathbf r_i}$.  We aim to recover $\Delta d({\mathbf r_i})$ by smoothing the individual logarithmic distance ratios $\Delta d_j$ for galaxy $j$.  Our smoothing algorithm is defined as 

\be
\Delta d({\mathbf r_i}) = {\sum_{j=1}^{N_j} 
  \Delta d_j\cos\theta_{i,j}\,e^{-rr_{i,j}/2}\,\sigma_j^{-3} 
  \over \sum_{j=1}^{N_j} e^{-rr_{i,j}/2}\,\sigma_j^{-3} }
\ee
where $\sigma_j$ is the smoothing length of the 3D Gaussian kernel for galaxy $j$, $\theta_{i,j}$ is the angle between the ${\mathbf r_i}$ for gridpoint $i$ and the ${\mathbf r_j}$ for galaxy $j$, and $rr_{i,j}$ is the square of the distance between gridpoint $i$ and galaxy $j$ in units of $\sigma_j$.  The summations in both the numerator and denominator of Equation 7 run over all $N_j$ galaxies for which $rr_{i,j}<9$.

We define the smoothing length $\sigma_j$ as a function of the fiducial kernel $\sigma^\prime = 15$ \mpc, weighted as a function of local density, $\delta_j$:
\be
\sigma_j=\sigma^\prime\left[\frac{\exp( \sum_{l=1}^{N}\ln\delta_l / N)}
           {\delta_j}\right]^\frac{1}{2}
\ee
where
\be 
\delta_j=\sum_{k=1}^{N_k} e^{-rr_{j,k}/2} 
\ee
and $rr_{j,k}$ is the square of the distance between galaxy $j$ and galaxy $k$ in units of $\sigma^\prime$.  The sum in Equation 8 is over all $N$ galaxies in the survey, while the sum in Equation 9 is over the $N_k$ galaxies within $3\sigma^\prime$ of galaxy $j$.  In this case, we set $\sigma^\prime = 15$ \mpc.  The mean smoothing length is $\langle\sigma_j\rangle=7.2$ \mpc, with a $1\sigma$ scatter of 4.2 \mpc.  The smallest smoothing length is 2.4 \mpc, while the largest is 29.0 \mpc.

The distribution of smoothing lengths only weakly depends on the fiducial value, $\sigma^\prime$.  The selection of the fiducial value is somewhat arbitrary, but we chose a value that we find clearly illustrate the broad trends in the velocity field.  In any case, the smoothed map is not used for any quantitative analysis in this paper.  All measurements of $\beta$ and the residual bulk flow use the unsmoothed $\Delta d$ values from the individual galaxies.

In Figures 4, 5, and 6, we show the adaptively smoothed velocity field along slices of SGX, SGY, and SGZ, for both the observed 2MTF dataset and the 2MRS and PSCz models.  SGX, SGY, and SGZ are the orthogonal axes in supergalactic cartesian coordinates.  They are defined by $SGX=r \cos (sgb) \cos (sgl)$, $SGY=r \cos (sgb) \sin (sgl)$, $SGZ=r \sin (sgb)$, where $r$ is the redshift space distance to the galaxy, and $sgl$ and $sgb$ are the supergalactic longitude and latitude respectively.

\begin{figure*} \centering 
\begin{minipage}{140mm}
\includegraphics[width=1.0\textwidth]{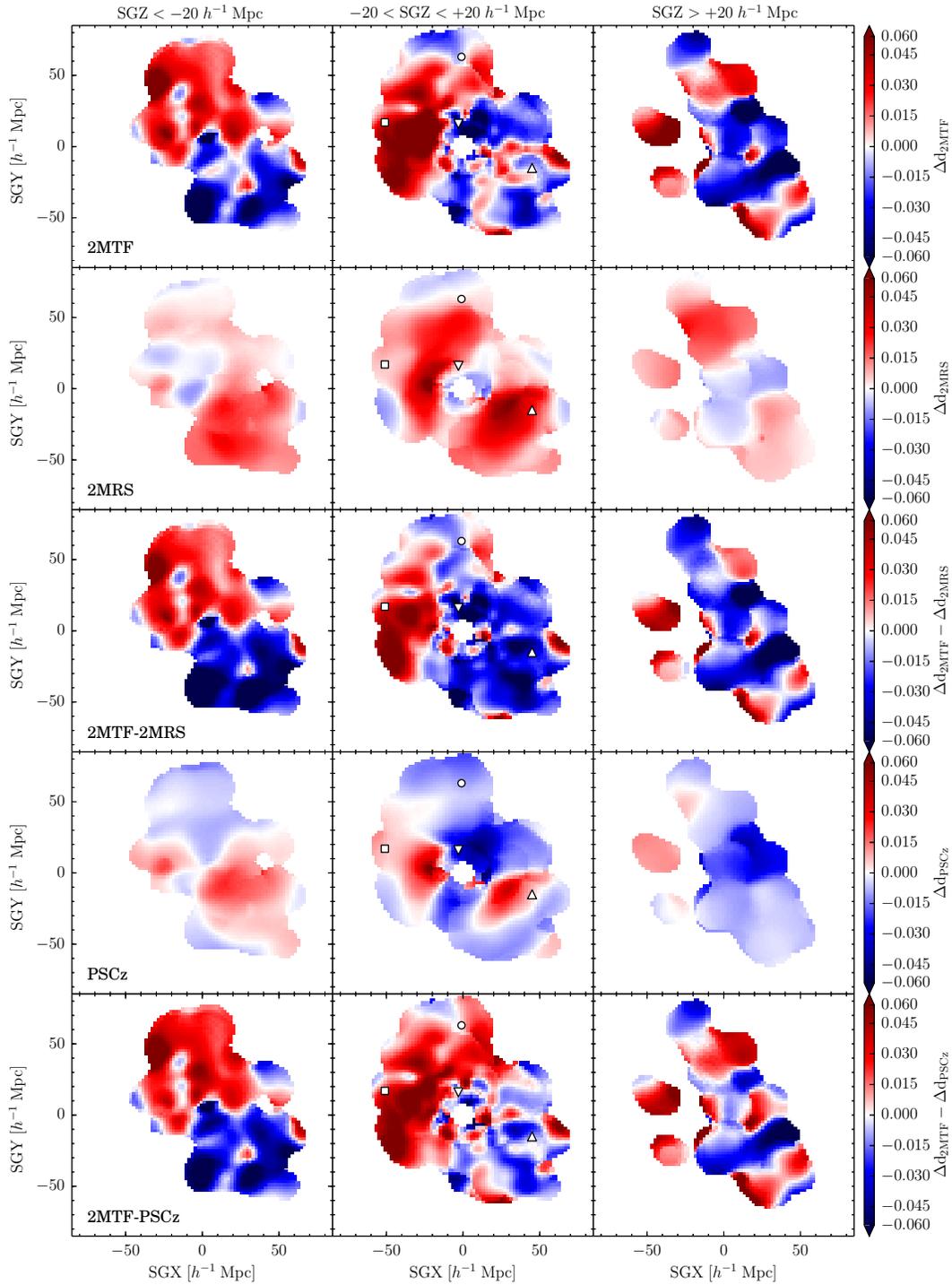} 
\caption{Adaptively smoothed maps of the nearby galaxy velocity field in supergalactic cartesian coordinates, in slices of $SGZ$.  In each case, the velocity field is given in logarithmic distance units ($\Delta d=\log(D_z / D_H)$, in the nomenclature of Section 3), as the logarithm of the ratio between the redshift distance and the true Hubble distance.  As shown in the colorbars for each panel, redder (bluer) colors correspond to more positive (negative) values of the logarithmic distance ratio, $\Delta d$, and thus more positive (negative) peculiar velocities.  The left column is the slice for $SGZ<-20$ \mpc.  The middle column is the $-20<SGZ<+20$ \mpc ~ slice.  The right column is the $SGZ>+20$ \mpc ~ slice.  The top row is the observed velocity field from 2MTF.  The second row is the 2MRS reconstructed velocity field, as derived by Erdo{\u g}du et al. (submitted).  The third row is the {\it difference} between the observed 2MTF $\Delta d$ values and the predicted 2MRS $\Delta d$ values.  That is, it is the residual in $\Delta d$ from 2MRS.  The fourth row is the PSCz reconstructed velocity field, as derived by \citet{branchini99}.  The fifth row is the 2MTF residual from PSCz.  Gridpoints are spaced 4 \mpc \ apart.  We also mark the approximate positions of several features of large scale structure: The Coma Cluster ($\bigcirc$), the Hydra-Centaurus Supercluster ($\square$), the Virgo Cluster ($\bigtriangledown$), and the Pisces-Perseus Supercluster ($\triangle$).\label{FIG4}} 
\end{minipage} 
\end{figure*}

\begin{figure*} \centering 
\begin{minipage}{140mm}
\includegraphics[width=1.0\textwidth]{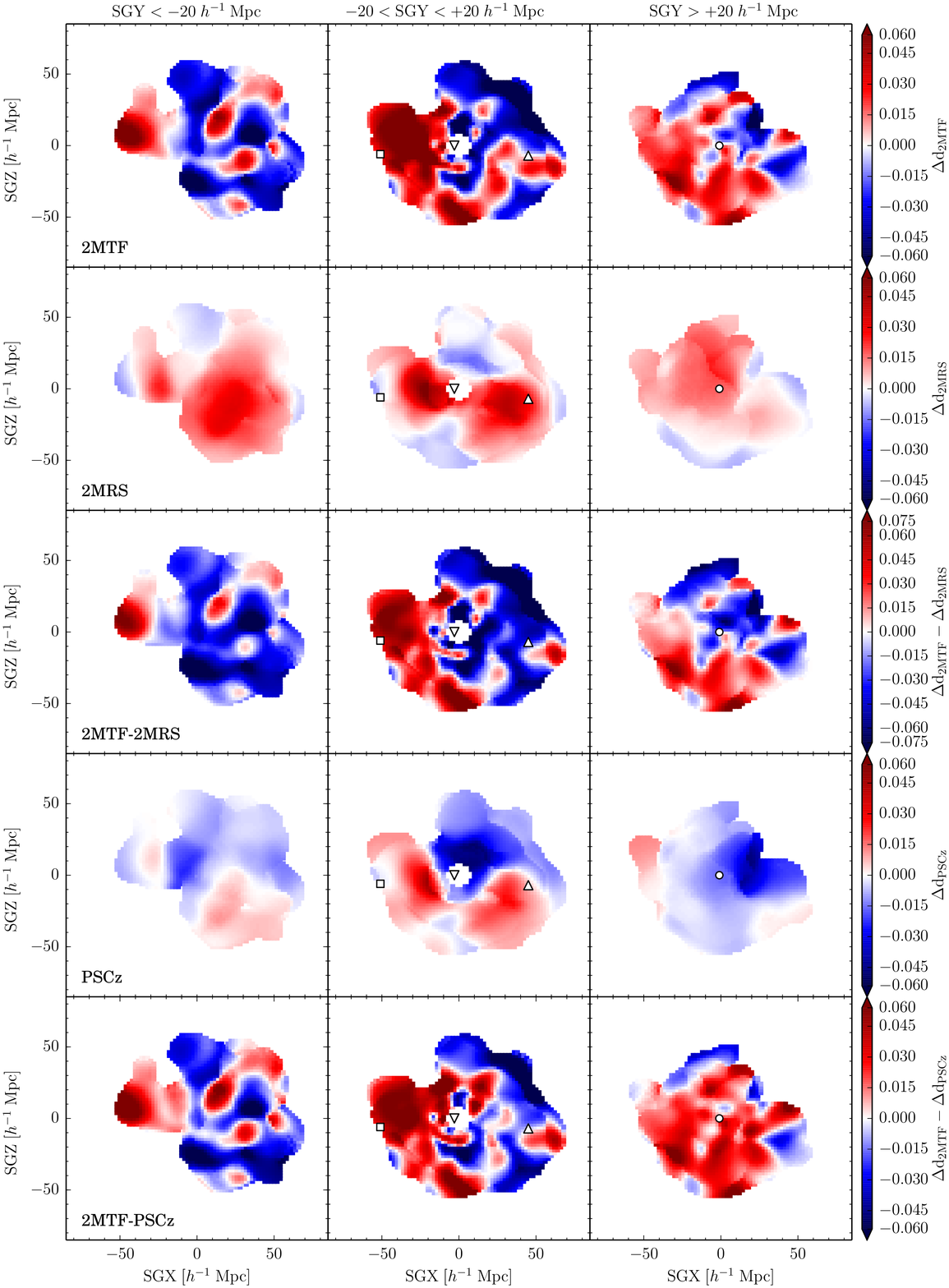} 
\caption{Same as Figure 4, but showing slices parallel to the SGY axis rather than the SGZ axis.\label{FIG5}} 
\end{minipage} 
\end{figure*}

\begin{figure*} \centering 
\begin{minipage}{140mm}
\includegraphics[width=1.0\textwidth]{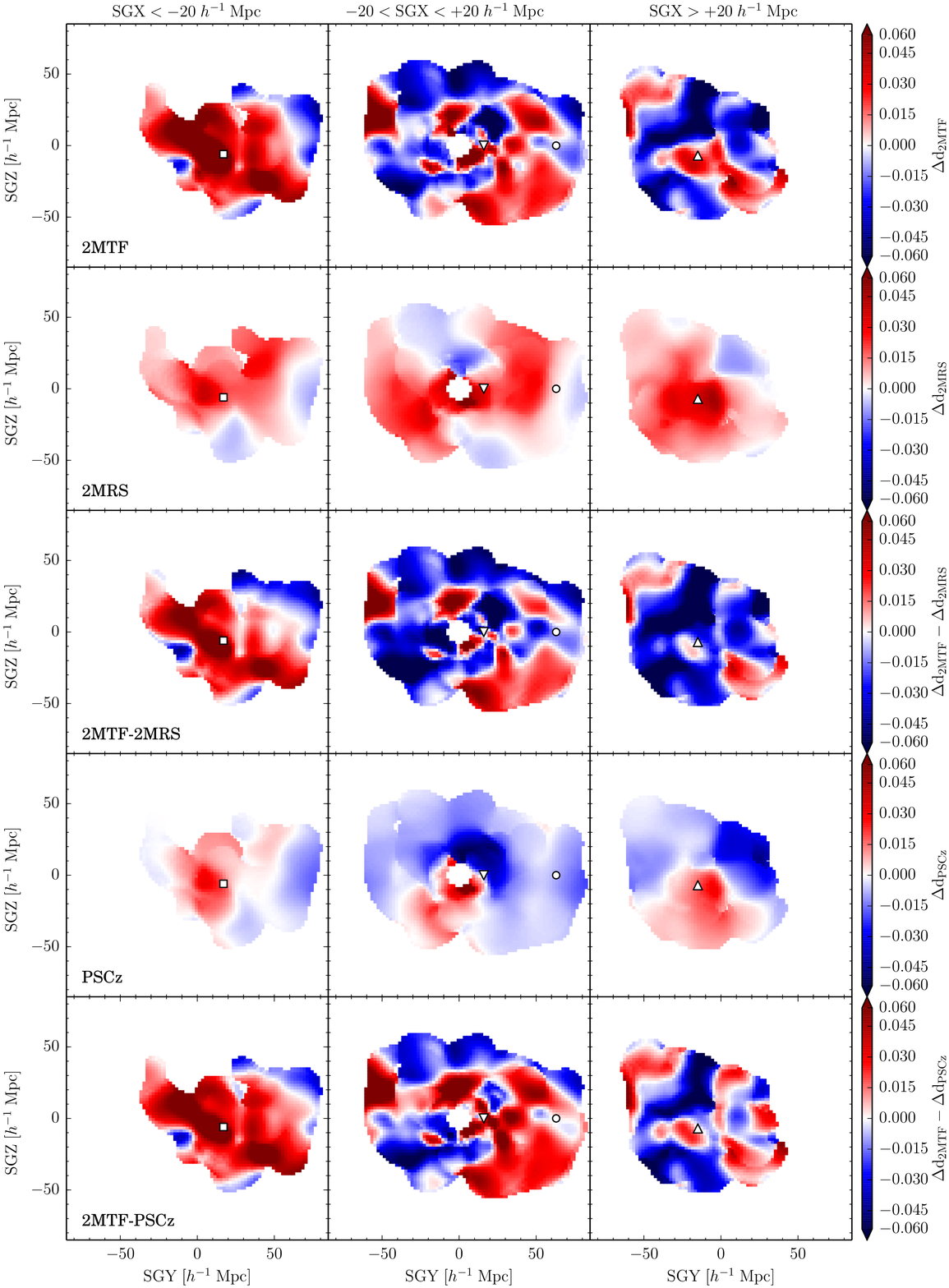} 
\caption{Same as Figure 4, but showing slices parallel to the SGX axis rather than the SGZ axis.\label{FIG6}} 
\end{minipage} 
\end{figure*}

In these figures, we also provide plots of the residual logarithmic distance ratios, $\Delta d_{2MTF} - \Delta d_{2MRS}$ and $\Delta d_{2MTF} - \Delta d_{PSCz}$.  These plots indicate features of the velocity field which are not predicted by the models.  We also include the approximate positions of various nearby features of large scale structure. (It should be noted that though objects such as the Hydra-Centaurus Supercluster and the Pisces-Perseus Supercluster are extended structures, we only display them as point sources so as not to obscure other features of the maps.)

The most notable features distinguishing the observed velocity field from the model velocity fields are: 1) A monopole deviation in $\Delta d$ between the 2MRS reconstruction and both the PSCz reconstruction and the 2MTF observation, with the former showing many more gridpoints with positive values of $\Delta d$, and 2) a dipole deviation between the observed velocity field and both model velocity fields.  Namely, the observed velocity field shows a noticeably large motion towards negative SGX and positive SGY, more or less in the direction of the Hydra-Centaurus and Shapley Superclusters.  \citet{hong14} measures this bulk flow at depths of 20, 30, and 40 \mpc\ with a Gaussian window function, finding values of $310.9\pm 33.9$, $280.8\pm 25.0$, and $292.3\pm 27.8$ \kms respectively.  While \citet{hong14} found these bulk flow values to be in good agreement with expectations from the $\Lambda$CDM model, we see here that the observed bulk flow does not appear to be replicated in the velocity field reconstructions.  We discuss this issue greater detail in Section 4.3.

Figures 4, 5, and 6 can also be directly compared to previous maps of the velocity field.  For example, \citet{theureau07} presents similar cosmographic plots of the nearby velocity field based on the Kinematics of the Local Universe survey (KLUN, \citealt{theureau05}), as do \citet{tully14} and \citet{hoffman15} using the Cosmic Flows 2 survey \citep{tully13}.  \citet{springob14} also includes very similar figures, as derived from the 6dF Galaxy Survey velocity field.  \citet{theureau07} Fig. 11 closely matches the `mid-plane' panel in Fig. 4 from this paper, for example.  In both cases, we see evidence of both foreground and background infall onto both the Pisces-Perseus Supercluster and Coma.  Both also show foreground infall towards Hydra-Centaurus, though our 2MTF cosmography plots are not deep enough to establish whether there is backside infall as seen by \citet{theureau07}.  Again, though, the largest apparent difference between the 2MTF and KLUN velocity field maps would appear to be the large dipole apparent in 2MTF, which is also notable in the 6dFGS cosmography, as described by \citet{springob14}.

\subsection{Fitting to the model velocity fields and measuring $\beta$}

As noted in Sections 2.2 and 2.3, the fiducial values of the redshift space distortion parameter $\beta$ are 0.4 for the 2MRS model and 0.5 for the PSCz model.  These are values that are assumed by the models, but we can measure the values for each model directly using 2MTF.  To clarify, $\beta$ is related to the matter density of the universe (in units of the critical density) $\Omega_m$ and the linear bias parameter $b$, according to
\be
\beta = \Omega_m^{0.55} / b
\ee
in a flat $\Lambda$CDM universe \citep{linder05}.  $b$ is defined as $b=\delta_g / \delta_m$, the ratio between overdensities in the galaxy density field and overdensities in the matter density field.  In the linear regime, the induced velocity $v$ at position $r$ is then
\be
v(r) = {\beta \over 4\pi} \int d^3 r^\prime {r^\prime - r \over |r^\prime - r|^3} \delta_g (r^\prime)
\ee
Since the amplitude of velocities is proportional to the value of $\beta$, the characteristic observed amplitude should scale with $\beta$.  Thus
\be
|v_{\mathrm{obs}}| / |v_{\mathrm{model}}| = \beta / \beta_{\mathrm{fid}}
\ee

Here, we employ a simple $\chi^2$ minimization in order to measure the value of $\beta$ for both the 2MRS and PSCz models, using the $N$ galaxies in the 2MTF sample:
\be
\chi_\nu^2 = \sum_{i=1}^N {(\Delta d_{i} - \Delta d_{\mathrm{model},i})^2 \over N\sigma_i^2}
\ee
where $\Delta d_{i}$ is the logarithmic distance ratio of galaxy $i$, and $\sigma_i$ is the uncertainty on this quantity.  $\Delta d_{model,i}$ is the logarithmic distance ratio for galaxy $i$ given by the 2MRS or PSCz model, with the model velocity scaled according to
\be
v_{i,j} / v_{i,\mathrm{fid}} = \beta_j / \beta_{\mathrm{fid}}
\ee
Here, we consider a range of possible $\beta$ values, $\beta_j$.  $v_{i,j}$ is then the model velocity for galaxy $i$, assuming $\beta$ value $\beta_j$.  The value of $\beta_j$ that minimizes $\chi_\nu^2$ according to Equation 13 is then our best-fit value of $\beta$.  The $68\%$ uncertainty on $\beta$ is then located at ${\chi_\nu}^2 = {\chi_{\nu , \mathrm{min}}}^2 (1\pm 1/N)$.

The resulting best-fit values of $\beta$ are $0.16\pm 0.04$ for 2MRS and $0.41\pm 0.04$ for PSCz.  The corresponding values of $\chi_\nu^2$ are 1.14 and 1.09 respectively.  These values can be compared to the values measured by other authors for the same velocity field reconstructions, as shown in Table 1.  As seen there, our measured value of $\beta$ for PSCz is on the low end of the literature values, though nearly identical to the value measured by \citet{branchini01}.  On the other hand, the value of $\beta$ for 2MRS is substantially lower than any of the literature values.

\begin{table*}
  \caption{The best fit values of $\beta$ measured for both the 2MRS and PSCz models, using $\chi^2$ minimization, for a variety of scenarios explained in greater detail in Section 4.2.  We also list several literature values for $\beta$ for each of the two models, again with 2MRS values in the left column and PSCz values in the right column.}
\label{tab:beta}
\begin{tabular}{lrrlrr}
\hline \hline
\multicolumn{1}{c}{2MRS} &
\multicolumn{1}{c}{} &
\multicolumn{1}{c}{} &
\multicolumn{1}{c}{PSCz} &
\multicolumn{1}{c}{} &
\multicolumn{1}{c}{} \\

\multicolumn{1}{c}{type of fit} &
\multicolumn{1}{c}{$\beta$} &
\multicolumn{1}{c}{$\chi_{\nu}^2$} &
\multicolumn{1}{c}{type of fit} &
\multicolumn{1}{c}{$\beta$} &
\multicolumn{1}{c}{$\chi_{\nu}^2$} \\
\hline

fiducial value & 0.40 &  & fiducial value & 0.50 & \\
standard fit & $0.17\pm 0.04$ & 1.15 & standard fit & $0.41\pm 0.04$ & 1.10 \\
fix $\langle\Delta d\rangle$ to 2MTF value & $0.31\pm 0.04$ & 1.10 & fix $\langle\Delta d\rangle$ to 2MTF value & $0.41\pm 0.04$ & 1.11 \\
fit $\langle\Delta d\rangle$ & $0.35\pm 0.04$ & 1.10 & fit $\langle\Delta d\rangle$ & $0.40\pm 0.05$ & 1.09 \\
\hline
{\bf Literature values} & & & & & \\
\citet{branchini12} & $0.32\pm 0.08$ & & \citet{branchini01} & $0.42\pm 0.04$ & \\
\citet{nusser12} & $0.32\pm 0.10$ & & \citet{nusser01} & $0.50\pm 0.10$ & \\
\citet{davis11a} & $0.33\pm 0.04$ & & \citet{zaroubi02} & $0.51\pm 0.06$ & \\
\citet{bilicki11} & $0.38\pm 0.04$ & & \citet{ma12} & $0.53\pm 0.01$ & \\
\citet{lavaux10} & $\sim 0.52$ & & \citet{turnbull12} & $0.53\pm 0.08$ & \\
\citet{pike05} & $0.55\pm 0.05$ & & \citet{radburn04} & $0.55\pm 0.06$ & \\
\hline
\end{tabular}
\end{table*}

As noted in Section 4.1, there is a significant monopole offset between 2MTF and PSCz.  We define $\langle\Delta d\rangle$ as the mean value of the logarithmic distance ratio $\Delta d$, averaged over galaxies in either the data or the models.  This gives us the zeropoint of the dataset in question.  We find that $\langle\Delta d\rangle = -0.005$ for 2MTF, $+0.013$ for the 2MRS model, and $-0.003$ for the PSCz model, if measured only at the positions of the 2MTF galaxies.  Thus, there is a significant monopole discrepancy between 2MTF and the 2MRS model.

In the case of both the data and the models, however, the zeropoint of the velocity field is based on a set of assumptions about the boundary conditions that cannot be independently tested with the data at hand.  The 2MRS model used here, for example, assumes zero gravitational potential at the survey boundary, though the authors also considered alternative sets of assumptions, such as zero net velocity at the survey boundary.  The 2MTF dataset likewise assumes $\langle\Delta d\rangle$ is $\sim 0$ in the survey volume by construction, because a monopole in the velocity field is completely degenerate with an offset in the zeropoint of the TF relation.  We thus consider how our measurement of $\beta$ might differ with a different set of assumptions about the zeropoint.  If we shift all of the $\Delta d$ values of 2MRS by 0.018 dex, so that its value of $\langle\Delta d\rangle$ matches that of 2MTF, then the best-fit value of $\beta$ is $0.31\pm 0.04$ (also listed in Table 1), representing better agreement with the literature values.  In Figure 7, we show this adaptively smoothed version of the 2MRS and PSCz reconstructions, with the $\langle\Delta d\rangle$ set to -0.005 dex for both models, as it is in 2MTF.

\begin{figure*} \centering 
\begin{minipage}{140mm}
\includegraphics[width=1.0\textwidth]{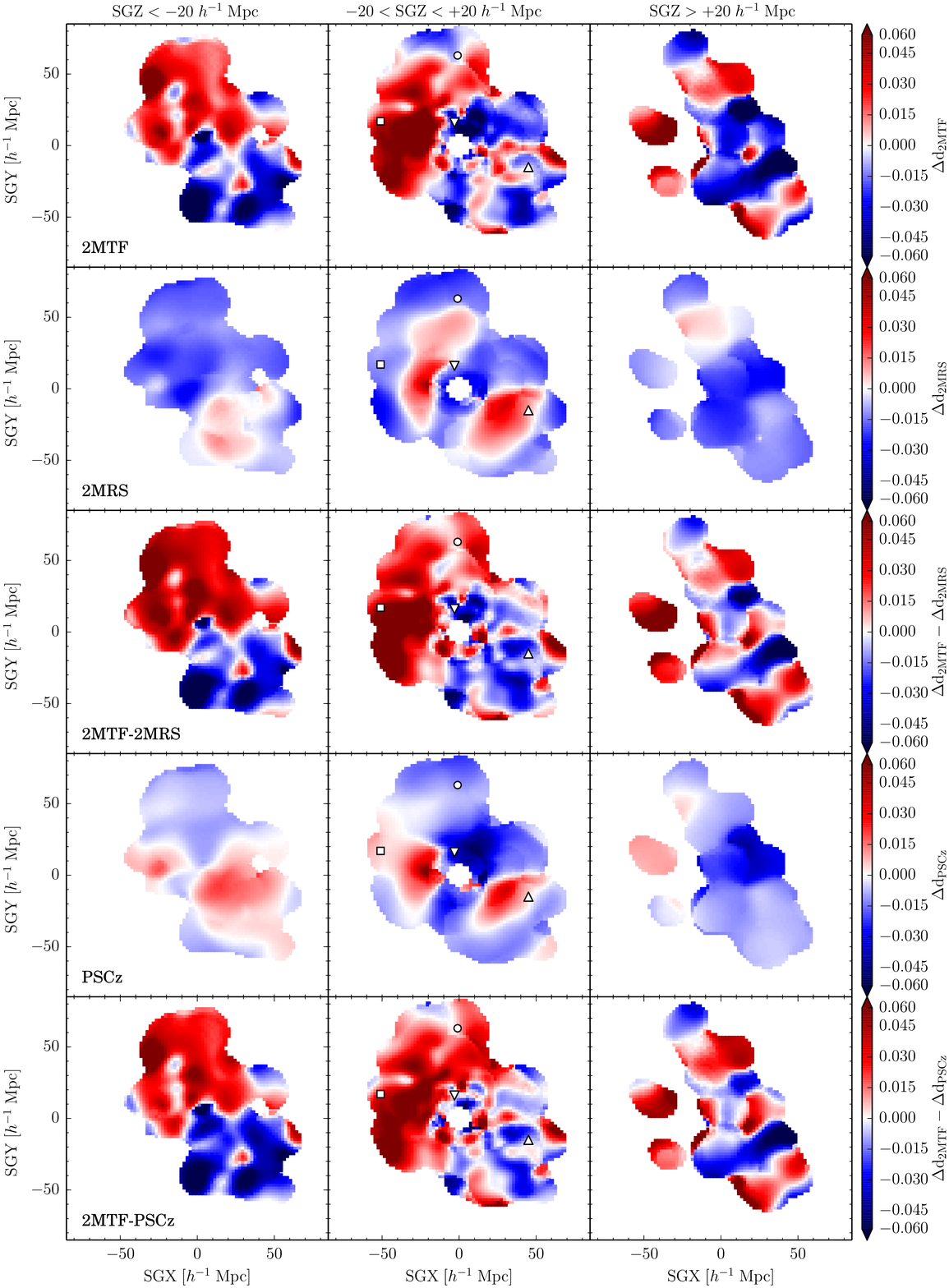} 
\caption{Same as Figure 4, but with the $\Delta d_{2MRS}$ and $\Delta d_{PSCz}$ values adjusted so that their mean value is -0.005 dex, which is the mean value of $\Delta d_{2MTF}$.\label{FIG7}} 
\end{minipage} 
\end{figure*}

Finally, we also allow the zeropoint of the models to float as a free parameter, adjusting the zeropoint to the value that gives us the minimum value of $\chi^2$.  The best fit value of $\langle\Delta d\rangle$ is $-0.010$ for 2MRS and $-0.014$ for PSCz.  This corresponds to $\beta=0.35\pm 0.04$ and $\beta=0.40\pm 0.05$ respectively.  This again shows better agreement with the literature values for 2MRS, but no impact on the $\beta$ value for PSCz.  The large monopole offset between 2MRS and 2MTF suggests that shifting $\langle\Delta d\rangle$, either by fixing it to the 2MTF value or fitting it via $\chi^2$ minimization, gives us a more meaningful estimate of $\beta$.  Hereafter, we refer to these two approaches as the `fix' and `fit' cases respectively.

Since $\beta$ is related to the bias parameter $b$ through the matter density $\Omega_m$ according to Equation 10, we can use either an independently measured value of $b$ to solve for $\Omega_m$ or an independently measured value of $\Omega_m$ to solve for $b$.  \citet{beutler12}, for example, using a prior on $H_0$ of $73.8\pm 2.4$ \kms Mpc$^{-1}$ as taken from \citet{riess11}, and derives values of $\Omega_m=0.250\pm 0.022$ and $b=1.48\pm 0.27$ for the redshift survey component of 6dFGS (\citealt{jones04}, \citealt{jones09}).  While we don't use 6dFGS results directly in this paper, the survey is responsible for the bulk of the Southern Hemisphere galaxies found in 2MRS.  We might then expect the surveys to have a similar value of the bias parameter.  If we assume this value of $b=1.48\pm 0.27$ for 2MRS, and use our measured $\beta=0.31\pm 0.04$  from the scenario in which we fix $\langle\Delta d\rangle$ to the 2MTF value, then that gives $\Omega_m = 0.24\pm 0.10$.  Whereas if we use our measured $\beta=0.35\pm 0.04$ from the case in which we fit $\langle\Delta d\rangle$ using $\chi^2$ minimization, then that gives $\Omega_m = 0.30\pm 0.12$.  Beutler also considers an $H_0$ prior of $67\pm 3.2$ \kms Mpc$^{-1}$ as taken from \citet{beutler11}, from which the authors derive $\Omega_m=0.279\pm 0.028$ and $b=1.52\pm 0.29$.  Using that value of $b$ would give us $\Omega_m=0.25\pm 0.10$ and $\Omega_m=0.32\pm 0.13$ in the `fix $\langle\Delta d\rangle$ to 2MTF value' and `fit $\langle\Delta d\rangle$' cases respectively.  Each of these estimates agrees with both the WMAP-9yr value ($\Omega_m=0.279\pm 0.025$, \citealt{bennett13}) and the Planck 2015 data release value ($\Omega_m=0.308\pm 0.012$, \citealt{planck15}).

We can also invert the problem, and assume the value of $\Omega_m$, to give us $b$.  The Planck value of $\Omega_m=0.308\pm 0.012$ \citep{planck15} gives us, for 2MRS, $b=1.69\pm 0.22$ and $b=1.49\pm 0.17$ in the `fixed' and `fit' cases respectively.  From this, we can derive the growth rate of structure $\Gamma \sim f \sigma_8$, where $\sigma_8$ refers to the amplitude of mass fluctuations on scales of 8 \mpc , while $f={\Omega_m}^{0.55}$.  This is related to $\sigma_{8,g}$, the amplitude of galaxy fluctuations on the same scale, by $\sigma_{8,g}=b\sigma_{8}$.  We can get the value of $\sigma_{8,g}$ for the 2MRS sample from \citet{westover07}, who measures $\sigma_{8,g}=0.97\pm 0.05$.  This then gives us $f\sigma_8 = \beta \sigma_{8,g} = 0.30\pm 0.04$ and $0.34\pm 0.04$ for the `fixed' and `fit' cases respectively.  These values are comparable to $f\sigma_8 = 0.31\pm 0.05$, as found by \citet{davis11a}, but somewhat lower than the $f\sigma_8 = 0.42\pm 0.07$ found by \citet{turnbull12}, the $f\sigma_8 = 0.42\pm 0.06$ found by \citet{beutler12}, and the $f\sigma_8 = 0.418\pm 0.065$ found by \citet{johnson14}.  The value also falls below both the WMAP-9yr value of $0.41\pm 0.02$ and the Planck 2015 data release value of $0.44\pm 0.01$.





\subsection{Residual bulk flow}

\citet{hong14} measured the bulk flow of the 2MTF sample using a $\chi^2$ minimization method.  The authors used the following relation for $\chi^2$:
\be
\chi^2 = \sum_{i=1}^N {(\Delta d_{i} - \Delta d_{\mathrm{model},i})^2 w_i^r w_i^d \over \sigma_i^2 \sum_{i=1}^N w_i^r w_i^d}
\ee
where $w_i^r$ is the weight assigned to galaxy $i$ based on the radial distribution of the sample, while $w_i^d$ is the weight that accounts for the variation in completeness as a function of declination.  Note that in this case, $\Delta d_{\mathrm{model},i}$ refers to a model in which the velocity field is characterized by a single dipole velocity vector ${\mathbf V_{\mathrm{bulk}}}$, which we refer to as the bulk flow.  This is in contrast to $\Delta d_{\mathrm{model},i}$ from Equation 13 in this paper, for which the model in question is the 2MRS or PSCz reconstruction.

The radial weight $w_i^r$ is set so that the redshift distribution of the sample is adjusted to match that of a Gaussian density profile, following \citet{watkins09}:
\be
\rho (r) \propto \exp(-r^2 / 2R_I^2 )
\ee
which translates to the number distribution
\be
n(r) \propto r^2 \exp(-r^2 / 2R_I^2 )
\ee
where $R_I$ is the characteristic depth of the bulk flow measurement.

The declination weight gives all galaxies north of $\delta=-40^{\circ}$ a weight of 1.00, and all galaxies south of $\delta=-40^{\circ}$ a weight of 2.08.  This is done to account for the fact that the 2MTF survey completeness is $\sim 2.08$ times greater north of $\delta=-40^{\circ}$ than it is south of that declination, owing to the differences in sensitivity of the northern and southern hemisphere telescopes used in the survey.

Errors on the bulk flow were estimated using a jackknife approach, as outlined in Section 4.1 of \citet{hong14}.  50 jackknife subsamples are created, with each randomly removing 2\% of the 2MTF sample.  The $\chi^2$ minimization is performed separately on each subsample, and the resulting scatter is then converted into a statistical uncertainty on the bulk flow, according to \citet{hong14} Equation 9.

The resulting bulk flow measurements at depths of 20\mpc, 30\mpc, and 40\mpc ~ are given in Table 1 of \citet{hong14}, and we reproduce those numbers in Table 2 of this paper.  As noted by \citet{hong14}, these values for the bulk flow are consistent at the $1\sigma$ level with the expectations given by the $\Lambda$CDM model (using the $\Lambda$CDM parameters $\Omega_m=0.27$, $\Omega_\Lambda=0.73$, $n_s=0.96$, as taken from the WMAP-7yr results of \citealt{larson11}, and the matter power spectrum generated by the CAMB package, as given by \citealt{lewis00}).

\begin{table*}
  \caption{Best fit values of the residual bulk flow, derived for Gaussian window functions of depth 20, 30, and 40 \mpc, using $\chi^2$ minimization.  In the top three rows, we show the total bulk flow as derived by \citet{hong14}.  The subsequent rows show residual bulk flows for the 2MRS and PSCz models under various assumptions about the best fit value of $\beta$, and the zeropoint ($\langle\Delta d\rangle$) of the velocity field for the model in question, as described in Section 4.2.  We provide the amplitude of the residual bulk flow $|{\mathbf V_{\mathrm{resid}}}|$, as well as its sky position, expressed in both spherical ($sgl$, $sgb$) and cartesian supergalactic ($V_{\mathrm{resid},sgx}$, $V_{\mathrm{resid},sgy}$, $V_{\mathrm{resid},sgz}$) coordinates.  We also translate the supergalactic coordinates into Galactic coordinates $l$ and $b$.  The bulk flow measurements from \citet{hong14} were made in Galactic coordinates, and so the measurement errors are only expressed in those coordinates.  On the other hand, the residual bulk flow measurements in this paper were made in supergalactic coordinates, and so errors are given only for that case.}
\label{tab:beta}
\begin{tabular}{lccccccccccc}
\hline \hline
\multicolumn{1}{c}{type of fit} &
\multicolumn{1}{c}{depth} &
\multicolumn{1}{c}{$\beta$} &
\multicolumn{1}{c}{$\langle\Delta d\rangle$} &
\multicolumn{1}{c}{$|{\mathbf V_{\mathrm{resid}}}|$} &
\multicolumn{1}{c}{$sgl$} &
\multicolumn{1}{c}{$sgb$} &
\multicolumn{1}{c}{$V_{\mathrm{resid},sgx}$} &
\multicolumn{1}{c}{$V_{\mathrm{resid},sgy}$} &
\multicolumn{1}{c}{$V_{\mathrm{resid},sgz}$} &
\multicolumn{1}{c}{$l$} &
\multicolumn{1}{c}{$b$} \\

\multicolumn{1}{c}{} &
\multicolumn{1}{c}{\mpc} &
\multicolumn{1}{c}{} &
\multicolumn{1}{c}{dex} &
\multicolumn{1}{c}{\kms} &
\multicolumn{1}{c}{deg} &
\multicolumn{1}{c}{deg} &
\multicolumn{1}{c}{\kms} &
\multicolumn{1}{c}{\kms} &
\multicolumn{1}{c}{\kms} &
\multicolumn{1}{c}{deg} &
\multicolumn{1}{c}{deg}  \\
\hline

bulk flow  & 20 & 0 & & $311\pm 34$ & 164 & -27 & & & & $288\pm 6$ & $11\pm 3$ \\
bulk flow & 30 & 0 & & $281\pm 25$ & 158 & -18 & & & & $296\pm 16$& $19\pm 6$ \\
bulk flow  & 40 & 0 & & $292\pm 28$ & 171 & -20 & & & & $296\pm 10$ & $6\pm 9$ \\
\hline
2MRS fiducial value & 20 & 0.40 & 0.013 & $228\pm 14$ & $169\pm 2$ & $-5\pm 4$ & $-223\pm 15$ & $43\pm 10$ & $-20\pm 15$ & 311 & 10 \\
2MRS fiducial value & 30 & 0.40 & 0.013 & $271\pm 14$ & $167\pm 2$ & $-12\pm 3$ & $-259\pm 13$ & $60\pm 12$ & $-55\pm 13$ & 304 & 11 \\
2MRS fiducial value & 40 & 0.40 & 0.013 & $282\pm 15$ & $165\pm 3$ & $-18\pm 3$ & $-258\pm 15$ & $71\pm 13$ & $-88\pm 13$ & 297 & 12 \\
2MRS standard fit & 20 & 0.16 & 0.013 & $304\pm 14$ & $167\pm 2$ & $-10\pm 3$ & $-292\pm 15$ & $70\pm 10$ & $-50\pm 16$ & 306 & 12 \\
2MRS standard fit & 30 & 0.16 & 0.013 & $330\pm 14$ & $166\pm 2$ & $-15\pm 2$ & $-309\pm 15$ & $77\pm 12$ & $-84\pm 12$ & 301 & 10 \\
2MRS standard fit & 40 & 0.16 & 0.013 & $330\pm 14$ & $165\pm 2$ & $-20\pm 2$ & $-299\pm 16$ & $80\pm 14$ & $-113\pm 11$ & 295 & 10 \\
2MRS fix $\langle\Delta d\rangle$ to 2MTF & 20 & 0.28 & -0.005 & $110\pm 17$ & $182\pm 6$ & $22\pm 10$ & $-102\pm 17$ & $-4\pm 10$ & $40\pm 16$ & 339 & 1 \\
2MRS fix $\langle\Delta d\rangle$ to 2MTF & 30 & 0.28 & -0.005 & $165\pm 18$ & $171\pm 4$ & $2\pm 5$ & $-163\pm 19$ & $27\pm 13$ & $4\pm 15$ & 318 & 9 \\
2MRS fix $\langle\Delta d\rangle$ to 2MTF & 40 & 0.28 & -0.005 & $189\pm 16$ & $165\pm 4$ & $-11\pm 4$ & $-179\pm 15$ & $49\pm 11$ & $-36\pm 14$ & 304 & 13 \\
2MRS fit $\langle\Delta d\rangle$ & 20 & 0.35 & -0.010 & $92\pm 20$ & $195\pm 10$ & $41\pm 10$ & $-67\pm 17$ & $-18\pm 10$ & $60\pm 17$ & 0 & -7 \\
2MRS fit $\langle\Delta d\rangle$ & 30 & 0.35 & -0.010 & $136\pm 18$ & $173\pm 5$ & $10\pm 7$ & $-133\pm 19$ & $16\pm 13$ & $23\pm 16$ & 327 & 8 \\
2MRS fit $\langle\Delta d\rangle$ & 40 & 0.35 & -0.010 & $159\pm 16$ & $165\pm 4$ & $-7\pm 5$ & $-153\pm 19$ & $41\pm 12$ & $-19\pm 14$ & 309 & 14 \\
\hline
PSCz fiducial value & 20 & 0.50 & -0.003 & $189\pm 20$ & $159\pm 6$ & $32\pm 7$ & $-150\pm 19$ & $57\pm 17$ & $100\pm 19$ & 349 & 21 \\
PSCz fiducial value & 30 & 0.50 & -0.003 & $216\pm 21$ & $156\pm 4$ & $10\pm 4$ & $-194\pm 20$ & $87\pm 15$ & $38\pm 13$ & 325 & 25 \\
PSCz fiducial value & 40 & 0.50 & -0.003 & $221\pm 17$ & $154\pm 4$ & $-2\pm 3$ & $-198\pm 15$ & $99\pm 17$ & $-6\pm 12$ & 312 & 26 \\
PSCz standard fit & 20 & 0.41 & -0.003 & $212\pm 18$ & $161\pm 5$ & $19\pm 5$ & $-189\pm 19$ & $65\pm 15$ & $69\pm 17$ & 335 & 20 \\
PSCz standard fit & 30 & 0.41 & -0.003 & $244\pm 20$ & $158\pm 3$ & $3\pm 3$ & $-226\pm 18$ & $90\pm 12$ & $12\pm 14$ & 329 & 23 \\
PSCz standard fit & 40 & 0.41 & -0.003 & $248\pm 16$ & $156\pm 3$ & $-7\pm 3$ & $-225\pm 15$ & $99\pm 14$ & $-29\pm 12$ & 307 & 23 \\
PSCz fix $\langle\Delta d\rangle$ to 2MTF & 20 & 0.41 & -0.005 & $211\pm 27$ & $54\pm 83$ & $86\pm 6$ & $8\pm 22$ & $11\pm 17$ & $210\pm 26$ & 50 & 10 \\
PSCz fix $\langle\Delta d\rangle$ to 2MTF & 30 & 0.41 & -0.005 & $147\pm 27$ & $138\pm 12$ & $62\pm 8$ & $-52\pm 23$ & $46\pm 13$ & $129\pm 21$ & 25 & 24 \\
PSCz fix $\langle\Delta d\rangle$ to 2MTF & 40 & 0.41 & -0.005 & $119\pm 26$ & $140\pm 10$ & $40\pm 11$ & $-70\pm 25$ & $59\pm 12$ & $76\pm 17$ & 2 & 34 \\
PSCz fit $\langle\Delta d\rangle$ & 20 & 0.40 & -0.014 & $218\pm 28$ & $44\pm 89$ & $86\pm 7$ & $10\pm 23$ & $10\pm 18$ & $217\pm 27$ & 50 & 9 \\
PSCz fit $\langle\Delta d\rangle$ & 30 & 0.40 & -0.014 & $150\pm 28$ & $138\pm 13$ & $63\pm 8$ & $-50\pm 23$ & $46\pm 14$ & $134\pm 21$ & 26 & 24 \\
PSCz fit $\langle\Delta d\rangle$ & 40 & 0.40 & -0.014 & $121\pm 27$ & $139\pm 10$ & $41\pm 12$ & $-69\pm 25$ & $59\pm 12$ & $79\pm 18$ & 4 & 34 \\
\hline
\end{tabular}
\end{table*}

Separate from the question of whether the bulk flow agrees with predictions from $\Lambda$CDM, there is also the question of whether the observed amplitude and direction of the bulk flow is predicted by the particular galaxy density distribution we observe in the local universe.  We now investigate a scenario in which the $\Delta d_{\mathrm{model},i}$ from Equation 15 is calculated using the model velocity of the 2MRS or PSCz models plus a residual bulk flow ${\mathbf V_{\mathrm{resid}}}$.  Fixing $\beta$ to the fiducial values of 0.40 and 0.50 respectively, and performing the $\chi^2$ minimization as in \citet{hong14}, we get residual bulk flow values of amplitude $\sim 200$ \kms at depths of 20, 30, and 40 \mpc\ for PSCz, with somewhat larger values for 2MRS.  

In addition to the fiducial values of $\beta$, we also calculate the residual bulk flow measured for the fitted values of $\beta$ using the methods described in Section 4.2.  Each of these values are listed in Table 2.  (Note that, for ease of comparison with previous papers on galaxy peculiar velocities and large scale structure, we have done the fitting in supergalactic coordinates, whereas \citealt{hong14} fit the bulk flow in Galactic coordinates.)  For both models, the amplitude of the residual bulk flow in the `standard fit' case is seen to be larger than in the fiducial case.  In fact, the 2MRS residual bulk flow for the standard fit of $\beta$ is even larger in amplitude than the total bulk flow.  As we noted in Section 4.2, though, we consider the `fix' and `fit' scenarios to offer more realistic estimates of the underlying value of $\beta$.  For both of these fits, for both the 2MRS and PSCz models, the residual bulk flow is $\sim 150$ \kms , which corresponds to roughly half of the total bulk flow.  This is also illustrated in Figure 8.

\begin{figure}
\includegraphics[width=0.9\columnwidth]{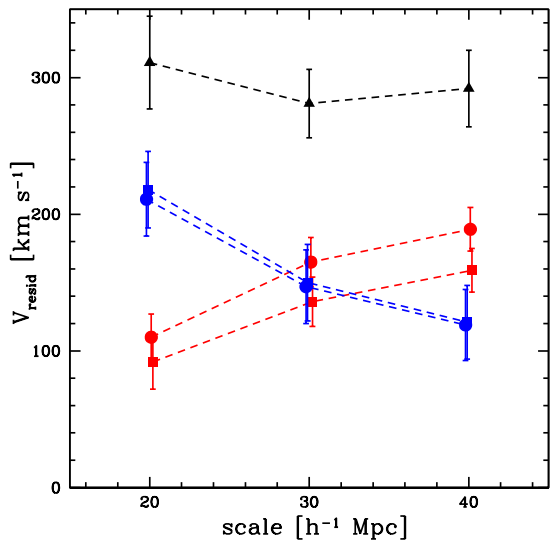}
\caption{Black triangles show the magnitude of the 2MTF bulk flow with Gaussian window functions at depths of 20, 30, and 40 \mpc , as derived by \citet{hong14}.  Red (blue) points represent residual bulk flow amplitudes at the same depths for the 2MRS (PSCz) model, with circles showing the values for the `fix $\langle\Delta d\rangle$ to 2MTF value' scenario, and squares for the `fit $\langle\Delta d\rangle$' scenario.  While each of the points lies exactly at either 20, 30, or 40 \mpc , we give a slight horizontal displacement to the points in order to make it more clear which set of errorbars corresponds to which datapoint.
\label{FIG8}}
\end{figure}

The residual bulk flow directions in both the `fiducial' and `standard fit' cases lie close to that of the {\it total} bulk flow direction.  In Figure 9, however, we show the residual bulk flow directions in both the `fix' and `fit' cases.  In both cases, the 2MRS residual extends from very low Galactic latitudes at the 20 \mpc\ scale to somewhat higher latitudes (and a direction very close to both the total bulk flow and the Hydra-Centaurus Supercluster) at the 40 \mpc\ scale.  The PSCz residual bulk flow in the `fix' and `fit' cases follow a similar pattern on the sky, but offset by $\sim 60^{\circ}$.  The PSCz residual at 40 \mpc\ is $\sim 45^{\circ}$ away from the Shapley Supercluster, and otherwise does not appear to be in the vicinity of any major features of large scale structure.

\begin{figure*}
\begin{minipage}{175mm}
\includegraphics[width=1.0\textwidth]{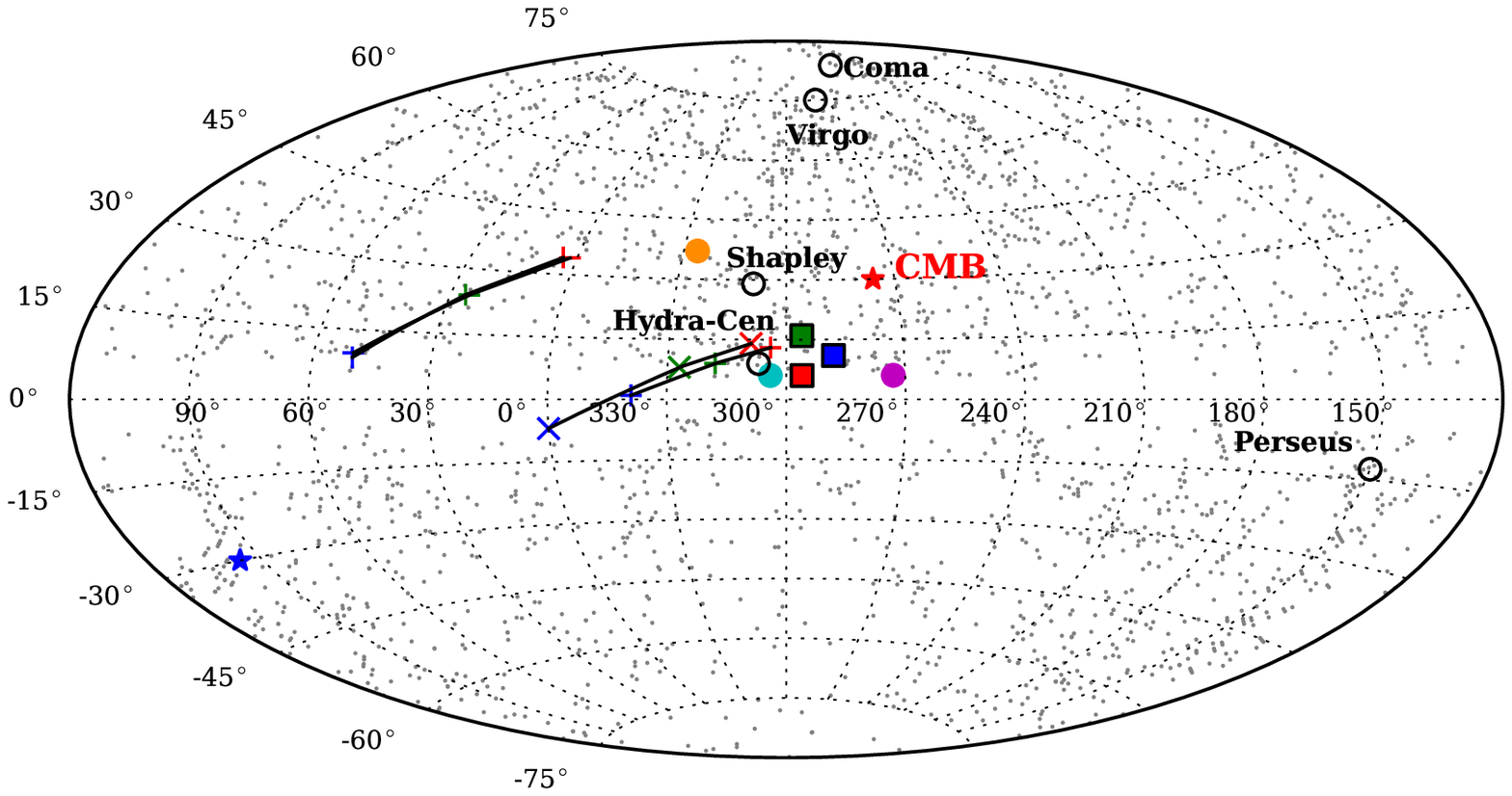}
\caption{The direction of the residual bulk flow of the 2MTF velocity field, as projected on the celestial sphere in Galactic coordinates, with respect to the 2MRS and PSCz models.  The residual bulk flow fit for 20, 30, and 40 \mpc ~ window functions are shown in blue, green, and red respectively, with black lines connecting them.  2MRS residual bulk flow shown as $+$'s for the `fix' case and $\times$'s for the `fit' case.  PSCz residual bulk flows are shown as $-$'s and $|$'s in the fix and fit cases respectively.  As also shown in Table 2, the fix and fit directions for PSCz are nearly identical.  We also show the total bulk flow as squares, coded with the same blue/green/red scheme for the 20, 30, and 40 \mpc ~ window functions.  The \citet{carrick15}, \citet{hudson04}, and \citet{magoulas12} residual bulk flows from the 2M++, PSCz, and 2MRS models respectively are shown as cyan, magenta, and orange circles.  The red star shows the location of the CMB dipole direction, with the blue star representing the anti-dipole direction.  Gray dots show the locations of 2MTF galaxies.  The approximate positions of various features of large scale structure are shown as open circles with labels.
\label{FIG9}}
\end{minipage}
\end{figure*}

Figures 4-7 also show the residual bulk flow cosmographically, in the $\Delta d_{2MTF} - \Delta d_{2MRS}$ and $\Delta d_{2MTF} - \Delta d_{PSCz}$ rows.  These can be compared to Figures 9 and 10 from \citet{springob14}, which show the cosmography of the 6dFGS peculiar velocity field, when the 2MRS and PSCz models are subtracted away from the observed $\Delta d$ values.  6dFGS is a Southern Hemisphere only survey, and is composed of galaxies whose mean redshift is more than twice as great as that of 2MTF.  Nonetheless, \citet{springob14} finds that the PSCz model offers a somewhat better fit to the 6dFGS velocity field than does the 2MRS model.  Cosmographically, \citet{springob14} Figures 9 and 10 show that the models underestimate both the outward flow towards Hydra-Centaurus and the Shapley Supercluster {\it and} the inward flow coming from a direction on nearly the opposite end of the sky, roughly coincident with the Cetus Supercluster.  In this paper, however, comparing the 2MTF velocity field to that of PSCz, we find that while the model underestimates the flow towards the Shapley Supercluster, there is not such a large underestimate of the inflow from the anti-Shapley direction.  (See the bottom central panel of Figure 4.)  Thus, at least judging from 2MTF, it seems possible that the deficiency of the models is that they underestimate the impact of known structures such as Shapley and Hydra-Centaurus, rather than that there are large structures outside the survey volume with a large influence on the local velocity field.

We can compare the residual bulk flow direction to results from other authors.  \citet{hudson04} compares peculiar velocities from the `Streaming Motions of Abell Clusters' (SMAC, \citealt{hudson01}) to the PSCz model, and finds a residual bulk flow of $372\pm 127$ \kms\ towards $(l, b) = (273^{\circ}, 6^{\circ})$, which is offset from our PSCz residual bulk flow by $\sim 90^{\circ}$.  \citet{magoulas12} compares 6dFGS peculiar velocities to the predictions of the 2MRS model, and finds a residual bulk flow of $273\pm 45$ towards $(l, b) = (326^{\circ}, 37^{\circ})$.  This is $\sim 30^{\circ}$ offset from our 2MRS residual bulk flow, but similarly close to the Shapley Supercluster.  \citet{carrick15} compares the SFI++ \citep{springob07} Tully-Fisher and First Amendment \citep{turnbull12} Type Ia Supernovae peculiar velocities to the predictions of the 2M++ reconstruction \citep{lavaux11}, finding a residual bulk flow of $159\pm 23$ \kms\ towards $(l, b) = (304^{\circ}, 6^{\circ})$ at 50\mpc , very close to our own measured 2MRS residual bulk flow at 40\mpc\ in both amplitude and direction.  There is heavy overlap between the 2MRS and 2M++ samples, so this agreement should not be surprising.

How does the amplitude of the residual bulk flow compare to our expectations, given a standard $\Lambda$CDM framework?  \citet{hudson04} examined the question of how consistent the rest frame of the PSCz gravity field should be with the CMB frame.  They estimated that while the contribution to the local bulk motion from sources beyond $\sim 200$ \mpc ~ should only be $\sim 50$ \kms , the contribution from systematic uncertainties ($\sim 90$ \kms ) and shot noise ($\sim 70$ \kms ) suggest a total uncertainty in PSCz's reconstruction of the bulk flow of $\sim 150$ \kms.  \citet{nusser14} reached a similar conclusion with regard to 2MRS.  Both \citet{davis11a} and \citet{carrick15} find agreement between the measured SFI++ bulk flow and the predictions of models roughly within this expected $\sim$150 \kms ~ range.  In this paper, we similarly find a similar residual bulk flow of $\sim 150$ \kms\ with respect to both the 2MRS and PSCz models, but only if one adjusts the comparison between data and model so as to remove the monopole deviation.  The amplitude of the residual bulk flow appears to be heavily dependent on assumptions about the boundary conditions of the model velocity field.

\section{Conclusions}

We have used adaptive kernel smoothing to present a cosmographic view of the local peculiar velocity field, as measured by the Tully-Fisher peculiar velocities from the 2MTF survey.  By extending all the way down to Galactic latitudes of $|b|=5^{\circ}$, 2MTF presents a more complete view of the velocity field at $cz<10,000$ \kms\ than has been seen before.

We compare the 2MTF velocity field to the reconstructed velocity field models from the 2MRS and PSCz redshift surveys.  We find best fit values of $\beta$ for these two models of $0.17\pm 0.04$ and $0.41\pm 0.04$ respectively, with $\chi_{\nu}^2$ values of 1.15 and 1.10 respectively.  There is a significant monopole offset between the 2MTF dataset and the 2MRS model.  As we have noted here, the zeropoint, and resulting monopole term in the velocity field, for both the data and models has been set on the basis of assumptions about the boundary conditions.  By changing those assumptions we may investigate the impact of the zeropoint on the measurment of $\beta$, as well as other parameters.  If we remove any monopole offset between data and models for both 2MRS and PSCz, then the best fit values of $\beta$ become $0.31\pm 0.04$ and $0.41\pm 0.04$, with $\chi_{\nu}^2$.  {\it Fitting} the zeropoint to minimize $\chi_{\nu}^2$ yields similar results.

These latter $\beta$ values are in line with previous estimates of $\beta$ for these same models from the literature.  They can be used in conjunction with other measurements to estimate parameters such as the matter density $\Omega_m$ and the growth rate of structure $f \sigma_8$.  We estimate values of $\Omega_m$ in line with previous measurements.  We also find a growth rate of structure of $\sim 0.3$, in line with the estimate by \citet{davis11a}, but somewhat lower than most other estimates.

\citet{hong14} used a $\chi^2$ minimization technique with Gaussian window function at depths of 20, 30, and 40 \mpc\ to measure the bulk flow of the 2MTF velocity field.  In this paper, we have now used that same technique to measure the {\it residual} bulk flow: the component of the bulk flow not accounted for by the velocity field models.  If we assume either the fiducial value of $\beta$ or the standard $\chi^2$ minimization fit, then the residual bulk flow is $\sim 200-300$ \kms, of roughly the same amplitude and direction as the total bulk flow.

However, when we adjust the zeropoint of the models, either by fixing them to the 2MTF value or fitting them via $\chi^2$ minimization, then the residual bulk flow drops to $\sim 150$ \kms.  This is in agreement with theoretical expectations, as discussed by \citet{hudson04} and \citet{nusser14}.  The residual bulk flow direction for 2MRS is at low Galactic latitude, and within $\sim 15^{\circ}$ of both Hydra-Centaurus and the total bulk flow direction.  On the other hand, the PSCz residual direction is at higher Galactic latitude ($34^{\circ}$ at the 40 \mpc\ scale) and not associated with any known features of large scale structure.  \citet{carrick15} finds a residual bulk flow comparing their TF and SNe peculiar velocity sample to the 2M++ sample which is strikingly similar in both amplitude and direction to our measured residual bulk flow for the 2MRS reconstruction.

This suggests that while 2MTF extends towards lower Galactic latitudes and creates a more complete all-sky sample, it does not reveal any significant influence on the velocity field from hidden structures not already apparent from peculiar velocity surveys with a larger zone of avoidance, like SFI++.  Both the total bulk flow and the residual bulk flow from the 2MRS model are at low Galactic latitude, but that was already observed in earlier surveys.  The inclusion of the lower Galactic latitude galaxies does not appear to create a substantial shift in the residual bulk flow direction.

\section*{Acknowledgements}

The authors gratefully acknowledge Martha Haynes, Riccardo
Giovanelli, and the ALFALFA team for supplying the latest ALFALFA
survey data.  We also thank Enzo Branchini for providing a copy of the PSCz reconstruction.

This research was conducted by the Australian Research Council Centre of Excellence for All-sky Astrophysics (CAASTRO), through project number CE110001020.

\end{document}